\documentstyle[pre,epsfig,aps,preprint]{revtex}

\def\beq{\begin{equation}}
\def\eeq{\end{equation}}

\def\eps_null{\varepsilon_0}
\def\half{\frac{1}{2}}
\def\vekt#1{\mbox{\boldmath $ #1  $}}

\def\pabl#1#2{\frac{\partial #1}{\partial #2}}

\def\Ehat{\hat{E}}
\def\Gammahat{\hat{\Gamma}}

\def\imagi{\mbox{i}}

\def\tret{t_{\mbox{\scriptsize ret}}}
\def\vektB{\vekt{B}}

\def\vektv{\vekt{v}}

\def\reff#1{(\ref{#1})}

\begin{document}


\title{Generation of ultra-short light pulses \\ by a rapidly ionizing thin foil}
\author{D.~Bauer, R.~R.~E.~Salomaa}
\address{Department of Engineering Physics and Mathematics, Helsinki University of Technology, P.O.Box 2200, FIN-02015 HUT, Finland}
\author{P.~Mulser}
\address{Theoretical Quantum Electronics (TQE), Darmstadt University of Technology, Hochschulstr. 4A,
D-64289 Darmstadt, Germany}

\date{\today}

\maketitle

\begin{abstract}
A thin and dense plasma layer is created when a sufficiently strong
laser pulse impinges on a solid target. The nonlinearity introduced by
the time-dependent electron density leads to the generation of
harmonics. The pulse duration of the harmonic radiation is related to
the risetime of the electron density and thus can be affected by the
shape of the incident pulse and its peak field strength. 
Results are presented from numerical particle-in-cell-simulations of
an intense laser pulse interacting with a thin foil target. 
An analytical model which shows how the harmonics are created is introduced. 
The proposed scheme might be a promising way towards the generation of
attosecond pulses.\\ 
PACS number(s): 52.40.Nk, 52.50.Jm, 52.65.Rr
\end{abstract}

\section{Introduction}
In recent years several mechanisms generating harmonics of
electromagnetic radiation have been discovered. 
Among these 
harmonics from gases \cite{gasharm}, harmonics from a laser pulse propagating
through underdense plasma \cite{underdense}, 
and harmonics production from the
plasma-vacuum boundary when a laser pulse impinges on a solid target
\cite{surface} are the most prominent ones. 
Especially the high-order
gas-harmonics, exhibiting a ``plateau'' instead of a rapid decrease
with the harmonic order, seem to be a promising source for xuv
``water-window''-radiation. 

Apart from the effort to make progress
towards shorter wavelengths, another goal is to achieve shorter pulse
durations because the temporal resolution in pump-probe experiments
clearly depends on the pulse length. One scheme proposed to generate
attosecond pulses is based on phase-matching pulse trains which are
produced by a laser pulse focused into a jet of rare gases
\cite{trains}. Another method makes use of the fact that the
efficiency of gas-harmonics generation is sensitive to the ellipticity
of the incident laser light \cite{ellipt}.

The method to generate an ultra-short {\em low order} harmonic laser
pulse as proposed in this paper is based on the time-dependent
electron density of the target material in laser pulse-solid
interaction.
The mechanism is thus entirely different from those mentioned above
which are based on phase-matching of nonlinear single atom-responses
\cite{trains}, relativistic (and thus nonlinear) electron trajectories
under the influence of the surrounding plasma \cite{underdense} or the oscillating vacuum-target-interface owing to
the $\vektv\times\vektB$-nonlinearity in the Lorentz force.
However, the key idea of the method proposed in this paper 
is similar to the one in Ref.\ \cite{ellipt},
namely to control the time duration of effective
harmonics production by the incident pulse itself. 
In Ref.\ \cite{ellipt} it
is the ellipticity which is the relevant parameter governing harmonics
creation while in the present paper it is the ionization rate which
determines the rise in the free electron density of the target. 
The harmonic pulse duration is of the order of this risetime and thus
can be tuned appropriately by varying the intensity of the incident pulse.

It turned out that although our method is capable of generating very
short pulses of third and fifth harmonic light (about 2 fundamental
laser cycles in length) the mechanism is not
appropriate for generating particularly {\em high order} harmonics
efficiently.  

In a recently published conference proceeding \cite{spanier} similar
studies of ``ionization harmonics'' are presented, but the authors did
not focus on the generation of ultra-short pulses there.

In section \ref{modeling} we introduce our simple one-dimensional
analytical model to study how the laser pulse propagation is influenced
by the ionizing thin foil. In section \ref{numresults} we discuss our
particle-in-cell simulation results. Finally, we
conclude in section \ref{concl}.

\section{One-dimensional modeling}\label{modeling}
We assume a linearly polarized laser pulse impinging perpendicularly on a thin foil target. In the following analytical and numerical calculations the whole setup is treated one dimensionally in space, i.e., the laser pulse propagates along $x$ and the electric field is in $y$-direction. The foil will be ionized by the pulse. To calculate the pulse propagation through a medium with varying free electron density one has to solve 
the inhomogeneous wave equation 
\beq \pabl{}{x^2} E(x,t) - \frac{1}{c^2} \pabl{}{t^2} E(x,t) = \frac{1}{\eps_null c^2} \pabl{}{t} j(x,t). \label{inh_wave} \eeq
The Green's function of this equation
is $G(x,x',t,t')=-c\Theta [ c(t-t')-\vert x-x'\vert ] / 2$
where $\Theta(y)$ is the step-function, i.e., $\Theta(y)=1$ for $y>0$ and 0  otherwise.
The solution of \reff{inh_wave} can be written as the sum of the incident field $E_0(x,t)$  and the radiation field produced by the current $j(x,t)$, i.e., $E(x,t)=E_0(x,t)+E_r(x,t)$,
with
\beq
E_r(x,t)=-\frac{1}{2c\eps_null}\int dt' \int dx'\,  \Theta[c(t-t')-\vert x-x'\vert ] \pabl{}{t'} j(x',t'). \label{radi}
\eeq
 
In order to model thin foils we now assume a delta-like current in space \cite{pegoraro}. If 
the thin foil is located at $x=0$ the current is
$j(x,t)=[-en_e(x,t) v_e(x,t)+Zen_i(x,t) v_i(x,t)] \ell \delta(x)$, 
with $n_{e,i}$ and $v_{e,i}$ the electron and ion density, and velocity, respectively, and $Z$ the ion's charge state.
Integrating $j(x,t)$ over $x$ one finds that the current per unit area
equals that of a ``real'' physical thin foil of thickness $\ell$ (as long as there is no strong electron density gradient across the foil).
Inserting the current $j(x,t)$ into \reff{radi} and performing
the spatial integration lead to
\beq E_r(x,t)= -\frac{\ell}{2c\eps_null } \int dt'\, \Theta[c(t-t')-\vert x\vert ] \ \pabl{}{t'} j_h(0,t') \label{Er_mediumstep}
\eeq
where $j_h(x,t)=-en_e(x,t) v_e(x,t)+Zen_i(x,t) v_i(x,t)$.
If we assume that the pulse hits the target at $t=0$ we finally get 
\beq 
E(x,t)=E_0(x,t)-\frac{\ell}{2c\eps_null} j_h(0,\tret) 
\label{eq_for_field} \eeq
for the electric field ($\tret=t-\vert x\vert /c$ is the retarded time).
The current $j_h(0,\tret)$ itself depends on the electric field. Neglecting
the ionic contribution to the current, we have
\beq j_h(0,\tret)=\frac{e^2}{m} n(0,\tret) \int_0^{\tret} dt' \, E(0,t') \eeq
where $n=n_e$.

Here it has been assumed that all newly created electrons are born with the
appropriate fluid element velocity and that collisional as well as relativistic effects are negligible. Besides, we neglect in our analytical treatment energy subtraction from the pulse due to the finite ionization energy of the target material. How this energy loss as well as momentum transfer due to
the velocity distribution  of the ionization produced electrons 
can be incorporated in a fluid description is studied in \cite{mulser}.   
All pulse intensities considered in this article do not cause relativistic electron motion. 

Supposing an ionization rate $ \Gamma $ applicable for pulse intensities under consideration has been chosen, the electron density $n$ in the foil is given by $n(0,\tret)=n_0 \left[ 1- \exp\left(-\int_0^{\tret} dt'\, \Gamma[ E(0,t') ]\right)\right]$.
When the target is fully ionized the electron density is $n_0$. We finally end up with the following integral equation for the electric field $E(x,t)$,
\beq E(x,t)=E_0(x,t)-\xi\left[ 1- \exp\left(-\int_0^{\tret} dt'\, \Gamma[ E(0,t') ]\right)\right] \int_0^{\tret} dt' \, E(0,t'), \label{integr_eq}  \eeq
where 
\beq
\xi= \frac{e^2 n_0 \ell}{2c\eps_null m}=\pi \left( \frac{\omega_p}{\omega_1}
\right)^2 \frac{\ell}{\lambda_1} \omega_1. \eeq
$\omega_p$ is the plasma frequency of the fully ionized target, $\omega_p^2=e^2 n_0/\eps_null m$, and $\omega_1$ and $\lambda_1$ are the incident EM wave's frequency and length, respectively. The dimensionless parameter $\xi/\omega_1$ determines how strong the propagation of the incident pulse is affected by the foil. For $\xi/\omega_1 \ll 1$ the foil is optically ``thin''.

If the foil is not pre-ionized or ionization is not completed already during the very early part of the pulse,  \reff{integr_eq} remains nonlinear due to the electron-density shape-factor
which depends on the electric field through the rate $\Gamma[E]$. Therefore
one expects harmonics in the transmitted and reflected light.

In what follows we will restrict ourselves to study \reff{integr_eq} in first order in $\xi/\omega_1$ (i.e., we assume a ``thin'' foil and iterate \reff{integr_eq} once).  At the position of the foil then 
\beq E(0,t)=E_0(0,t)-\xi\left[ 1- \exp\left(-\int_0^{t} dt'\, \Gamma[ E_0(0,t') ]\right)\right] \int_0^{t} dt' \, E_0(0,t') \label{integr_eq_first}  \eeq
holds. Here, the difficulty is to calculate  $\exp(-\int_0^{t} dt'\, \Gamma[ E_0(0,t') ])$. The ionization rate $\Gamma$ depends on the {\em absolute value} of the electric field, i.e., the rate has two maxima per fundamental laser cycle. Supposed that the pulse envelope $\Ehat_0$ is sufficiently adiabatic the rate may be expanded in a Fourier-series with even multiples of the
fundamental frequency only, and a slowly time-dependent envelope $\Gammahat$,
\beq \Gamma[\vert E_0(t)\vert ] = \Gammahat \left\{ \half a_0  + \sum_{n=1}^{\infty} ( a_{2n} \cos 2n\omega_1 t + b_{2n} \sin 2n\omega_1 t ) \right\} .\label{fourier_expans}  \eeq

Since the rate $\Gamma$ is a complicated functional of the field, in general all terms in the expansion \reff{fourier_expans} are present. However, if we
assume the incident pulse (divided by its amplitude) to be an even function in time, $E_0(t) \sim \cos \omega_1 t$,
 all coefficients $b_{2n}$ in \reff{fourier_expans} vanish. 
This finally leads to
\begin{eqnarray}
E(t) &=& \Ehat_0 \cos \omega_1 t \label{harmonics_result_i} \\
&& - \frac{\xi}{\omega_1} \left\{ 1-\exp(-\alpha_0 t) \prod_{n=1}^{\infty} \sum_{m=-\infty}^{\infty} (-\imagi)^m \exp(-\imagi 2 n m \omega_1 t ) I_m(\alpha_n) \right\} \nonumber \\
&& \qquad \times \Ehat_0 \sin\omega_1 t . \nonumber
\end{eqnarray} 
Here, $\alpha_0=\Gammahat a_0/2$, $\alpha_n=\Gammahat a_{2n}/2n\omega_1$, and $I_m$ is the modified Bessel-function. Note that $\alpha_0, \alpha_n$ are slowly time-dependent due to their dependence on $\Gammahat$. 

From \reff{harmonics_result_i} we can deduce that by ionization in first order of $\xi/\omega_1$ only {\em odd harmonics} will be produced: In \reff{harmonics_result_i} the term in curly brackets  is composed of even harmonics, but multiplied with $\sin\omega_1 t$ odd harmonics are created. 

Harmonics production is enhanced when $\xi/\omega_1$ is
increased. Therefore one expects that increasing the density or the
thickness of the foil acts in favor of the creation of
harmonics. However, one has to bear in mind that a perturbative
treatment with respect to $\xi/\omega_1$ preceded here. Furthermore,
increasing the density $n_0$ or the thickness $\ell$ simply makes the foil less transparent.

\section{Numerical results} \label{numresults}
A 1D2V particle-in-cell (PIC) code was used to simulate laser pulse-solid interaction. In order to incorporate ionization the rate
equations governing the ionization state of the target were solved
during each ``PIC-cycle'' \cite{bird}. For simplicity only one
ionization state with the ionization energy of hydrogen (13.6~eV) was
assumed. Landau's tunneling-rate \cite{landau} was used which is a
reasonable choice for the field strengths and frequencies under
consideration. The short risetime of the electron density forces a
tiny time step. Usually one fundamental laser cycle (wavelength) was
sampled by 1000 temporal (spatial) gridpoints. About $10^4$ computer
particles, sampling the physical charge densities of the thin foil
were found to be sufficient. The ions were mobile (although this is
unimportant for the effect under consideration) and 1836 times heavier
than the electrons (hydrogen).  

In Fig.~\ref{figure_one} numerically computed spectra of the transmitted light are shown for 5 different peak field strengths $\Ehat_0$, corresponding to intensities $I=4.0,4.8,6.5\times 10^{14}$ and $1.1,1.6\times 10^{15}$~W/cm$^2$. All other parameters were held constant: wavelength $\lambda_1=815$~nm, foil thickness $\ell=\lambda_1/10$,  incident $\sin^2$-shaped laser pulse of duration $T=30$~fs, and the density was the critical one with respect to the fundamental frequency, i.e., $n_0=n_c=1.68\times 10^{21}$~cm$^{-3}$. 

The higher the field strength, the broader are the harmonics peaks in
the spectrum. 
The pulse length of the harmonics radiation is closely 
related to the risetime of the electron density in the foil since as
soon as the density remains constant harmonics production will stop.  
In Fig.\ \ref{figure_two} the normalized electron density is plotted
vs time for the 5 field strengths of Fig.\ \ref{figure_one}. A
risetime covering $3$ fundamental periods for the weakest pulse and
only one cycle for the strongest pulse can be inferred from the plot
(each stair in the density corresponds to one half cycle).   
The density risetime is very sensitive to field strength and pulse
shape. Increasing the field strength leads to a decreasing density
risetime and hence to a shorter harmonics pulse length.  
However, the conversion efficiency decreases when the field strength
of the incident pulse is increased since the nonlinearity switches off
too soon. Furthermore the ``harmonics'' peaks are shifted and
asymmetrically broadened if  complete ionization occurs within only
one  fundamental cycle (or even less).  
In the limit of a step-like behavior of the electron density the
spectrum resembles  the Fourier-transform of the $\Theta$-function
with no $\omega_1$-harmonics structure at all.   

In Fig.~\ref{figure_one} the pulse length of the harmonic radiation
can be estimated by fitting the peaks in the spectrum to a Fourier
transformed ``test envelope'' $\sim\sin^2\pi t/T_n$. $T_n$ is the
pulse duration of the $n$th harmonic. For the 5 cases of Fig.\
\ref{figure_one} one finds for the pulse length of the 3rd harmonic
$T_3=3.3,3.0,2.3,2.0,1.9$ times the fundamental period
$\tau=2\pi/\omega_1$. A lower limit for $T_3$ certainly is $\tau$
itself because  a shorter risetime of the electron density leads to a
vanishing $\omega_1$-structure in the spectrum. The power in the 3rd
harmonic is about $10^{-6}$ of the fundamental. This conversion
efficiency is similar to the one in Ref.\ \cite{underdense} while all
other methods mentioned in the introduction are superior as far as power transferred
to (short pulse) harmonic radiation is concerned.   

One may object that the incident pulse intensity we took in our
numerical simulations was already small (at
least for ``up-to-date'' short pulse laser systems) so that the
3rd-harmonic pulse with only a millionth of its intensity is not 
acceptable at all. 
However, the incident pulse might be a stronger but defocused
pulse so that the 3rd harmonic output, when focused, becomes
considerable. Besides, using a shorter fundamental wavelength (and
correspondingly a thinner and/or denser foil) would require a higher
field strength to fully ionize the target within the same number of
cycles. By examining  the dimensionless parameters $\xi/\omega_1$ and
$\Gammahat/\omega_1$ one can estimate the ``experimental parameters''
$\ell$, $n_0$ and $\Ehat$ in order to meet the desired harmonic pulse
duration $T_3$. The practical limit for $T_3$ found in the numerical
simulations is about  $2\tau$. If the incident light has already a
rather short wavelength (e.g., if light, produced by one of the
high-harmonics mechanisms described above are used) then the
$2\tau$-limit can be shifted towards the attosecond-domain. However,
in that case it would be certainly a challenging task to find the
optimal parameters $\ell$ and $n_0$ for a  manufacturable thin foil.  
In recent experiments with thin foil targets \cite{pisa,jena} the thickness $\ell$ was
about $70$--$100$~nm and the electron density $n_0$ (when fully
ionized) was of the order of $5\times 10^{23}$~cm$^{-3}$, i.e., by a
factor 100 denser than assumed in our numerical simulations presented
so far.

One may wonder if it is really essential to take a {\em thin}
foil. When a sufficiently strong laser pulse impinges on a thick
overdense target only a thin plasma layer is created anyway and our
analytical treatment in section \ref{modeling} should therefore apply
as well. We thus expect also from thick targets short harmonic
radiation, at least in reflection.

In Figure \ref{spek_thick}a spectra calculated from timerows  of
the transverse electric field $E(x,t)$ in front (reflected light), in the
middle, and at the rear edge (transmitted light) 
of a thicker target are presented.
The target was one wave length thick, $\ell = \lambda$, and ten times
overcritical, $n_0=10n_c=1.68\times 10^{22}$~cm$^{-3}$.
The incident pulse had an intensity $I=1.6\times
10^{15}$~W/cm$^2$. The other parameters were $\lambda_1=815$~nm, $
T=30$~fs, square-sine in shape again.
The spectra have a rather complicated structure owing to the complex
spatial and temporal dependence of the free electron density. This can
be seen in Figure \ref{spek_thick}b where the target density is shown
as a contour plot vs time and space. During the pulse the target gets
fully ionized only in a thin layer at the front. The electron density
quickly rolls off and at the rear side not even the critical density
is reached. In the spectra we clearly observe the well-known effect
that higher frequency radiation penetrates easier the target than
light of the fundamental frequency does since for third harmonic light the target density is only slightly
overcritical while it is even undercritical for the 5th harmonic. This
effect was utilized as an experimental method to determine the plasma
density \cite{jena}. We observe also frequency shifts since
the peaks are not located any longer at the odd harmonics' positions. 
Obviously, thick targets are not as easily accessible by means of 
simple physical reasoning as it is the case for thin foils where the
$\delta$-current model described in section \ref{modeling} works well.

Our experiences from several PIC runs with different target densities
and thicknesses can be summarized as follows: as soon as there is a
density gradient across the foil the spectra of the reflected and
transmitted light get distorted (compared to the thin foil-spectra in
Fig.\ \ref{figure_one}). To avoid this density gradient the optimal
$\ell \times n$-product is about $0.1\lambda_1 n_c$
which is, with present day technology, not easy to achieve. Using a
higher incident pulse intensity in order to ionize the target more
homogeneously reduces the density risetime and thus the conversion
efficiency.

Apart from these ``technical problems'' one may argue that during the plasma formation process 
electron-ion-collisions might be important, especially for the
relatively low field strengths about $\approx 10^{15}$~W/cm$^2$. Since
it is during the plasma formation where the harmonics are produced
there might be serious distortions in the spectrum of the transmitted
light. In order to take electron-ion-collisions into account we
introduced a collision frequency $\nu_{ei}$ into our 1D2V PIC-code. This
leads to dissipation of energy due to friction of the oscillating
charge-sheets (note that in a 1D PIC-code each ``computer-particle''
represents an actual charge sheet \cite{bird}). The dissipated energy
is used to determine a ``sheet-temperature'' which, in turn, enters
into $\nu_{ei}$. We found collisions causing mainly distortions at
high frequencies but the 3rd and 5th harmonic-peaks were almost
unaltered. For stronger incident pulses the effect of collisions is
even less.  

With our PIC-code we also examined the effects of  energy subtraction
due to the ionization energy of the target material (according to the
model in \cite{mulser}). The most prominent effect, as far as
harmonics generation is concerned, is that ionization gets slowed down
slightly (which can be compensated by choosing a higher incident pulse
intensity).

It is worth mentioning that the observed effect of harmonics
production due to the rise of the electron density in a thin foil may be
used to {\em measure} the ionization time of the foil instead of
presupposing an ionization rate. This would offer an opportunity to check
the validity of ionization models experimentally.

\section{Conclusion} \label{concl}
In summary, we have studied the spectrum of a perpendicularly incident
laser pulse when transmitted through a rapidly ionizing foil.
A simple analytical method was utilized to show what the underlying
mechanism for the generation of short pulse odd harmonic radiation is.
The pulse duration of the harmonic radiation
is only a few cycles with respect to the frequency of the incident
laser light. The pulse length is governed by the risetime of the
electron density in the target and therefore it can be easily tuned
through adjusting the peak field strength of the incident pulse. This
might be a promising way towards the generation of attosecond pulses.

\section*{Acknowledgment}
This work was supported by the European Commission through the TMR
Network SILASI (Super Intense Laser Pulse-Solid Interaction), No.\
ERBFMRX-CT96-0043. One of the authors (D.B.) would like to thank Mr.\
G.\ Baier for stimulating discussions.


\begin{figure}
\caption{\label{figure_one} Spectra of the transmitted light for 5
different peak field strengths $\Ehat_0$, corresponding to intensities
$I=4.0\times 10^{14}$ (solid), $4.8,6.5\times 10^{14}$, $1.1\times
10^{15}$ (dotted), $1.6\times 10^{15}$~W/cm$^2$ (solid again). 
All other parameters were held constant: wavelength $\lambda_1=815$~nm, foil thickness $\ell=\lambda_1/10$, $T=30$~fs incident $\sin^2$-shaped laser pulse, and density $n_0=n_c=1.68\times 10^{21}$~cm$^{-3}$.  
The higher the field strength the broader are the harmonics peaks in the spectrum. }
\end{figure}

\begin{figure}
\caption{\label{figure_two} Normalized electron density vs time for the 5 field strengths of Fig.\ \ref{figure_one}. A risetime covering $3$ fundamental periods for the weakest pulse and only one cycle for the strongest pulse can be inferred (each stair in the density corresponds to one half cycle).}
\end{figure}

\begin{figure}
\caption{\label{spek_thick} (a) The spectra taken in front (solid), in
the middle (dotted), and at the rear side (dashed) of the
$\ell=\lambda_1$-foil, normalized to the fundamental peak of the
reflected light. The incident pulse had an intensity $I=1.6\times
10^{15}$~W/cm$^2$. The other parameters were $\lambda_1=815$~nm, $
T=30$~fs, square-sine in shape. For the fundamental frequency, the
plasma layer at the front end is 10 times overcritical while higher frequency light can
penetrate more easily. Compared to the clear harmonics structure in Fig.\
\ref{figure_one} the spectra are rather distorted. The ionization
harmonics peaks appear up-shifted. (b) The target electron density vs
space (scaled in fundamental wave length) and time (in fundamental
cycles). At the front end the target is (for the
incident light) 10 times overcritical while at the rear side the
target is not fully ionized ($n=n_0/100=n_c/10$).} 
\end{figure}

\end{document}